

\magnification=\magstep1
\baselineskip=16pt
\hfuzz=6pt
$ $

\vskip 1in

\centerline{\bf Quantum algorithms for supervised and unsupervised
machine learning}

\vskip 1cm

\centerline{\it Seth Lloyd$^{1,3}$, Masoud Mohseni$^2$, Patrick Rebentrost$^1$}

\centerline{1. Massachusetts Institute of Technology, Research Laboratory
for Electronics}

\centerline{2. Google Research}  
\centerline{ 3. To whom correspondence
should be addressed: slloyd@mit.edu}

\vskip 1cm
\noindent{\it Abstract:}  Machine-learning tasks frequently
involve problems of manipulating and classifying large
numbers of vectors in high-dimensional spaces.
Classical algorithms for solving such problems
typically take time polynomial in the number of vectors and
the dimension of the space.  Quantum computers are good at
manipulating high-dimensional vectors in large tensor product
spaces.  This paper provides supervised and unsupervised
quantum machine learning algorithms for cluster assignment and
cluster finding.  Quantum machine learning can take time
logarithmic in both the number of vectors and their dimension,
an exponential speed-up over classical algorithms.

\vskip 1cm

In machine learning, information processors perform tasks of sorting,
assembling, assimilating, and classifying information [1-2].  In supervised
learning, the machine infers a function from a set of training examples.
In unsupervised learning the machine tries to find hidden structure in
unlabeled data. Recent studies and applications 
focus in particular on the problem of large-scale machine learning [2] --
big data -- where the training set and/or the number of features is large. 
Various results on quantum 
machine learning investigate the use of quantum information processors
to perform machine learning tasks [3-9], including pattern matching [3],
Probably Approximately Correct learning [4], feedback learning for
quantum measurement [5], binary classifiers [6-7], and quantum 
support vector machines [8].  

This paper shows that quantum machine learning
can provide exponential speed-ups over classical computers for
a variety of learning tasks.  The intuition is straightforward. 
Machine learning is about manipulating and classifying large amounts
of data.  The data is typically post-processed and 
ordered in arrays (vectors) and arrays of arrays 
(tensor products): quantum computers are good at manipulating vectors 
and tensor products in high-dimensional
spaces.   In different machine learning settings,
the speed-up plays out in different fashions.   
First, classical data expressed in the form
of $N$-dimensional complex vectors can be mapped onto
a quantum states over $\log_2 N$ qubits: when the data
is stored in a quantum random access memory (qRAM), this
mapping takes $O(\log_2 N)$ steps [10-16].  
Once it is in quantum form, the data can be post-processed by
various quantum algorithms (quantum Fourier transforms [17],
matrix inversion [18], etc.),  
which take time $O({\rm poly}(\log N))$.
Estimating distances and inner products between post-processed 
vectors in $N$-dimensional 
vector spaces then takes time $O(\log N)$ on a quantum computer.
By contrast, as noted by Aaronson [19], sampling 
and estimating distances and inner products between post-processed 
vectors on a classical computer is apparently exponentially hard.
Quantum machine learning provides an exponential speed-ups over all
known classical algorithms for problems involving evaluating
distances and inner products between large vectors. 

In this paper, we show that the problem of assigning $N$-dimensional vectors 
to one of several clusters of $M$ states takes time $O(\log( MN))$ on 
a quantum computer, compared with time $O({\rm poly}(MN))$
for the best known classical algorithm.   That is, quantum machine
learning can provide an exponential speed-up for problems involving
large numbers of vectors as well (``big quantum data").
We present a quantum version of Lloyd's algorithm for 
performing $k$-means clustering: using a novel version of the
quantum adiabatic algorithm one can classify $M$ vectors into $k$ clusters in
time $O(k\log kMN)$.

Finally, we note that in addition to supplying exponential speed-ups
in both number of vectors and their dimension, quantum machine
learning allows enhanced privacy: only $O(\log(MN))$ calls
to the quantum 
data-base are required to perform cluster assignment, while 
$O(MN)$ are required to uncover the actual data.  The data-base user
can still obtain information about the desired patterns, while the
data-base owner is assured that the user has only accessed
an exponentially small fraction of the data base.

\smallskip\noindent{\it Data preparation and pre-processing}

Classically, data sets are typically presented as arrays of symbols
and numbers.  We assume that our data sets that consist
of arrays of numbers (vectors), and arrays of arrays (collections
of vectors), originally stored in random access memory (RAM)
in the classical case, or in quantum random access memory (qRAM)
in the quantum case [10-16].  The key feature of quantum machine
learning is that quantum random access memory allows us to access
the data in quantum parallel.  Begin with state preparation.
Consider the vector $N = 2^n$
dimensional complex vector $\vec v$
with components $\{ v_i = |v_i|e^{i\phi_i} \}$.
Assume that $\{|v_i|, \phi_i\}$ are stored as floating point
numbers in quantum random access memory.  
Constructing the $\log_2 N$ qubit quantum state 
$|v\rangle = |\vec v|^{-1/2} \vec v$ then takes $O(\log_2 N)$ steps
as long as the sub-norms $n_\ell = \sum_{i=1}^\ell |v_i|^2$ can
be estimated efficiently [20-22].  Alternatively, we can assume that
these sub-norms are also given in qRAM in which case any quantum
state can be constructed in $O(\log N)$ steps.

Once the exponentially compressed quantum versions of the vectors have
been created, we can postprocess them using quantum Fourier transforms,
matrix inversion, etc., to create vectors
of the form $QFT |v\rangle$, $f(A) |v\rangle$, where $A$ is a sparse
Hermitian matrix and $f$ is a computable function, e.g., $f(A) = A^{-1}$.  
The postprocessing takes time $O({\rm poly}(\log N) )$
As will now be shown, this allows us to evaluate generalized
inner products $\langle u|QFT |v\rangle$, and $\langle u| f(A)|v\rangle$
between the quantum vectors.  By contrast, as noted by Aaronson [19], the
best known algorithms for
evaluating the classical versions of these generalized inner products
$\vec u^\dagger FT \vec v$, $\vec u^\dagger f(A) \vec v$, 
via sampling and classical postprocessing takes time $O({\rm poly} N)$.

\smallskip\noindent{\it Supervised cluster assignment:} 

Consider the task of assigning a post-processed vector 
$\vec u \in R^N$ to one of two sets $V,W$,
given $M$ representative samples $\vec v_j \in V$ and  $M$ samples $\vec w_k \in W$.  A common method for such an assignment
is to evaluate the distance $|\vec u - (1/M) \sum_j \vec v_j |$ 
between $\vec u$ and the mean of the vectors in $V$, 
and to assign $\vec u$ to $V$ if this
distance  is smaller than the distance between $\vec u$ and the mean
of $W$.  We now exhibit a quantum algorithm for performing the
assignment problem that takes time $O( \epsilon^{-1}\log(MN) )$.
In the quantum assignment problem, assume that the vectors are presented
as quantum states $|u\rangle$, $\{|v_j\rangle\}$, $\{|w_k\rangle\}$.
If the vectors are are not normalized to one, assume that their
normalizations $|\vec v_j|$, $|\vec w_k|$ are given separately.

To evaluate the distance from $\vec u$ to the mean
of $V$, adjoin an ancilla variable with $M + 1$ states.
First, construct the state $| \psi \rangle
= (1/\sqrt 2) \big( |0\rangle|u\rangle
+ (1/\sqrt{M}) \sum_{j=1}^{M} |j\rangle|v_j\rangle \big)$ for
system and ancilla by querying the quantum RAM or by the
subroutine described above.  
Second, use a swap test [17] to perform a projective
measurement on the ancilla alone to see if it is in the state 
$|\phi\rangle = (1/\sqrt{Z}) ( |\vec u| ~ |0\rangle 
- (1/\sqrt M) \sum_j |\vec v_j| ~ |j\rangle )$ for the
ancilla alone, where $Z = |\vec u|^2 + (1/M) \sum_j |\vec v_j|^2$.  
It is straightforward to verify that the desired distance, 
$ | ~ \vec u - (1/M) \sum_j  \vec v_j ~ |^2$,
is equal to $Z$ times the probability of success for this measurement.

The state $|\phi\rangle$ can be generated by using
quantum access to the norms
together with quantum simulation to apply the unitary
transformation $e^{-iHt}$, where
$ H = (~|\vec u| |0\rangle \langle 0| + \sum_j |\vec v_j| |j\rangle
\langle j|~) \otimes \sigma_x$, to the state
$(1/\sqrt 2) (~ |0\rangle - (1/\sqrt M) \sum_j |j\rangle~)\otimes |0\rangle$.
The result is the state
$$\eqalign{
&(1/\sqrt 2) \big(~ \cos(|\vec u| t) ~ |0\rangle - (1/\sqrt M) \sum_j
\cos(|\vec v_j|t) ~ |j\rangle~ \big)\otimes |0\rangle \cr
& -(i/\sqrt 2) \big( ~ \sin(|\vec u|t) ~ |0\rangle -
(1/\sqrt M) \sum_j \sin(|\vec v_j|t) ~ |j\rangle ~ \big)\otimes |1\rangle.}
\eqno(1)$$
Choosing $t$ so that $|\vec u| t, |\vec v_j|t \ll 1$
and measuring the ancilla bit then yields the state $|\phi\rangle$
with probability $(1/2) (|\vec u|^2 + (1/M) \sum_j |\vec v_j|^2)t^2
= Z^2 t^2$.   This procedure creates the desired state and, when
repeated, also allows the quantity $Z$ to be estimated.  
A more efficient way to create
the state and to estimate $Z$ to accuracy $\epsilon$ is to use
Grover's algorithm/quantum counting [17].  Quantum 
counting takes time $O(\epsilon^{-1} \log M)$, and also allows
quantum coherence to be preserved during the state creation.

\smallskip\noindent{\it Unsupervised quantum learning:}

The exponential quantum speed-up above holds for supervised
learning.  A similar speed-up extends to unsupervised learning.
Consider the $k$-means problem of assigning $M$ vectors to
$k$ clusters in a way that minimizes the average distance
to the centroid of the cluster.  The standard method for solving
$k$-means is Lloyd's algorithm [1-2] (no relation to the co-author
of this paper): (0) choose the initial centroid randomly or by a
method such as $k$-means$^{++}$; (1) 
assign each vector to the cluster with
the closest mean;  (2) re-calculate the centroids of
the clusters; repeat steps (1-2) until a stationary assignment is attained.
When classical estimation of the distance to the centroids in
the $N$-dimensional space takes time $O(N)$, each
step of the classical
algorithm takes time $O(M^2 N)$, while the
quantum Lloyd's algorithm takes time $O(M\log(MN))$.
The additional factor of $M$ in both classical and quantum
algorithms arises because every vector is tested individually
for reassignment at each step.

The quantum Lloyd's algorithm can be improved by
noting that the $k$-means problem can be rephrased as a quadratic
programming problem which is amenable to solution by the adiabatic
algorithm.  As will now be seen, such unsupervised quantum machine
learning takes time at most $O(k\log(MN))$ and can even
take only $O(\log(kMN) )$.  In order to reduce the dependence
on the number of vectors from $O(M\log M)$ to $O(\log M)$,
the output of the computation can no longer be a list
of the $M$ vectors and their cluster assignments.  Instead,
the output is a quantum state 
$|\chi\rangle = (1/\sqrt M)\sum_{j}  |c_j\rangle |j\rangle 
= (1/\sqrt M)\sum_{c,j\in c}  |c\rangle |j\rangle
$
that contains the labels $j$ of vectors 
correlated with 
their cluster assignments $c_j$ in superposition: we can then sample
from that state to obtain a statistical picture of the clustering.
The procedure for constructing the clustering state $|\chi\rangle$
via the quantum adiabatic algorithm is given in the supplementary material.  
The algorithm takes time no greater than
 $O(\epsilon^{-1} k \log kMN)$ to construct this
state to accuracy $\epsilon$, and could take time
as little as $O(\epsilon^{-1} \log kMN)$ if the clusters are relatively
well separated, so that the gap of the adiabatic algorithm
is $O(1)$.  

Any algorithm that reveals the assignment of all $M$ vectors necessarily
takes time $O(M)$ merely to print out the output.  Many questions
about the $k$-means clustering can be answered using smaller outputs.
As we now show, adiabatic algorithms provide a powerful method
for answering clustering questions.
First, look at the problem of finding initial seeds for the clusters.
As the efficiency of the $k$-means$^{++}$ algorithm shows,
the performance of Lloyd's algorithm, classical or quantum, depends
strongly on a good choice of initial seeds.  Initial seed vectors
should be spread as far apart from each other as possible.
Begin the adiabatic seed-finding algorithm in the state
$ |\Psi\rangle = |\psi\rangle_1 \otimes \ldots |\psi\rangle_k$, where
$|\psi\rangle = (1/\sqrt M) \sum_{j = 1}^M |j\rangle$
is the uniform superposition of vector labels,
and with initial Hamiltonian $ H_0 = 1-|\Psi\rangle\langle \Psi|$.

The distance-finding algorithm given above allows us to apply any
Hamiltonian of the form,
$$H_s = \sum_{j_1\ldots j_k} f( \{ |\vec v_{j_\ell} - \vec v_{j_{\ell'}}|^2 \})
|j_1\rangle\langle j_1| \otimes \ldots \otimes |j_k\rangle \langle j_k|.
\eqno(2)$$ 
To find good seeds for $k$-means, use a final Hamiltonian for
the adiabatic algorithm of the form (2) with
$f = - \sum_{\ell, \ell' =1}^k |\vec v_{j_\ell} - \vec v_{j_{\ell'}}|^2$.
The ground state of this
final Hamiltonian is the seed set that maximizes the average
distance squared between seeds.  

We can also use the adiabatic algorithm to find sets of $r$ vectors
that should lie in the same cluster.  Here the final Hamiltonian
is of the form  
$$H_c = \sum_{j_1\ldots j_r} f( \{ |\vec v_{j_\ell} - \vec v_{j_{\ell'}}|^2 \})
|j_1\rangle\langle j_1| \otimes \ldots \otimes |j_r\rangle \langle j_r|,
\eqno(3)$$
where
$f =  \sum_{\ell, \ell' = 1}^r |\vec v_{j_\ell} - \vec v_{j_{\ell'}}|^2
+  \kappa\delta_{j_\ell,j_{\ell'}}$, $\kappa > 0$.  
Because of the overall positive
sign, the distance term now rewards sets of vectors that are 
clustered closely, while the  $\kappa\delta_{j_\ell,j_{\ell'}}$ ensures
that the vectors in the $\ell$ and $\ell'$ positions are different
(we already know that a vector lies in the same cluster as itself).
To find such sets of vectors that are expected to lie in the
same cluster can take time $O( r \log MN )$, depending on 
the probability of success of the quantum adiabatic 
algorithm (see next paragraph).
Combining this `attractive' Hamiltonian with the `repulsive' Hamiltonian
of (2) allows one to find $kr$ representative groups of $r$ vectors from
each of the $k$ clusters.

The success of the quantum adiabatic algorithm in finding the ground
state of the final Hamiltonian relies on traversing the minimum gap
point of the quantum phase transition between the initial and final
Hamiltonians sufficiently slowly.  Finding the optimal seed set of size
$k$ classically is a combinatorially hard problem in $k$, and finding
the optimal cluster of $r$ vectors is combinatorially hard in $r$.
Accordingly, the minimum gap and time
to find the ground state may well scale exponentially in $k,r$.
Indeed, optimal $k$-means is an NP-complete problem which we do
not expect to solve in polynomial time on a computer, classical or
quantum.  Approximate solutions of these hard problems are well
within the grasp of the quantum adiabatic algorithm, however.
$k$-means$^{++}$ does not require an optimal seed set, but merely
a good seed set with well-separated vectors.  In addition, in 
$k$-means we are interested in finding various sets of highly clustered
vectors, not only the optimal set.  Even running the algorithm
for a time linear in $O(k\log MN)$ is likely to suffice to construct
pretty good seed sets and clusters.  
We can reasonably hope that an adiabatic quantum computer
that traverses the minimum gap in finite time $\tau$ at
finite temperature $T$ should be able to find approximate solutions
whose energy is within $\max\{O(kT), O(\hbar/\tau)\}$ of the minimum energy.
The question of how well
the adiabatic algorithm performs on average is an open one.

\smallskip\noindent{\it Extension to nonlinear metrics:}

The quantum algorithm for determining distance can be generalized
to nonlinear metrics to compare vectors using the results of [18].  
Given $q$ copies
of $|u\rangle$, $|v\rangle$, the quantum phase algorithm
can be used to evaluate $(\langle u| \langle v|)^{\otimes k}
L (|u\rangle|v\rangle)^{\otimes k}$ for arbitrary Hermitian $L$, 
allowing distance measures that are $q$'th order polynomials in 
the $u_j$, $v_j$.  Measurement of the expectation value of
$L$ to accuracy $\epsilon$ using quantum counting
requires $O( \epsilon^{-1} q\log N)$ steps.  Once again,
the quantum algorithm reduces the dimension dependence of the
evaluation of distance to $O(\log N)$.    

\smallskip\noindent{\it Discussion:}

The power of quantum computers to manipulate large numbers
of high-dimensional vectors
makes them natural systems for performing vector-based machine learning
tasks.  Operations that involve taking vector
dot products, overlaps, norms, etc., in $N$-dimensional
vector spaces that take time $O(N)$ in the
classical machine learning algorithms, take time $O(\log N)$ in
the quantum version.   These abilities, combined with the quantum
linear systems algorithm [18], represent a powerful suite of
tools for manipulating large amounts of data.
Once the data has been processed in a quantum form, as in
the adiabatic quantum algorithm for search engine ranking [23],
then measurements can be made on the processed data to reveal
aspects of the data that can take exponentially longer to
reveal by classical algorithms
Here, we presented a quantum algorithm
for assigning a vector to clusters of $M$ vectors that takes
time $O(\log MN)$, an exponential speed-up in both $M$ (quantum
big data) and $N$.  We used this algorithm as a subroutine for the standard
$k$-means algorithm to provide an exponential speed-up for unsupervised learning
(quantum Lloyd's algorithm) via the adiabatic algorithm.  

Currently, the rate of generation of electronic data generated per year
is estimated to be on the order of $10^{18}$ bits.  This entire
data set could be represented by a quantum state using $60$ bits,
and the clustering analysis could be performed using a few
hundred operations.  Even if the number of bits to be analyzed
were to expand to the entire information content of the universe
within the particle horizon, $ O(10^{90}\approx 2^{300})$ bits, 
in principle the data representation and analysis would be well 
within the capacity of a relatively small quantum computer.

The generic nature of the quantum speed-ups for dealing with
large numbers of high-dimensional vectors suggests that a wide
variety of machine learning algorithms may be susceptible 
to exponential speed-up on a quantum computer.
Quantum machine learning also provides advantages in terms
of privacy: the data base itself is of size $O(MN)$, but 
the owner of the data base supplies only $O(\log MN)$ quantum
bits to the user who is performing the quantum machine learning
algorithm.   In addition to supplying an exponential speed-up
over classical machine learning algorithms, quantum machine
learning methods for analyzing large data sets (`big
quantum data') supply significant advantages in terms of
privacy for the owners of that data.

\vfill 
\noindent{\it Acknowledgments:} This work was supported by DARPA,
Google, NSF, ARO under a MURI program, Jeffrey Epstein, and FQXi.
The authors thank Scott Aaronson for helpful discussions.

\vfil\eject

\noindent{\it References:}

\vskip 1cm

\smallskip\noindent [1] E. Alpaydın {\it 
Introduction to Machine Learning 
(Adaptive Computation and Machine Learning)}, MIT Press, Cambridge (2004).

\smallskip\noindent [2] D. Mackay {\it Information Theory, Inference and 
Learning Algorithms,} Cambridge University Press, Cambridge (2003).

\smallskip\noindent [3] M. Sasaki, A. Carlini, 
{\it Phys. Rev. A} {\bf 66}, 022303 (2002); arXiv: quant-ph/0202173.

\smallskip\noindent [4]
R. A. Servedio and S. J. Gortler, 
{\it SIAM J. Comput.} {\bf 33}, 1067, (2004);
arXiv: quant-ph/0007036.

\smallskip\noindent [5]
A. Hentschel, B.C. Sanders, {\it Phys. Rev. Lett.}{ \bf 104} (2010), 
063603; arXiv: 0910.0762.

\smallskip\noindent [6] H. Neven, V.S. Denchev, G.  Rose, 
W.G. Macready,
Training a Large Scale Classifier with the Quantum Adiabatic Algorithm,
arXiv: quant-ph/0811.0416; arXiv: 0912.0779.

\smallskip\noindent [7] K.L. Pudenz, D.A. Lidar, 
 {\it Quant. Inf. Proc. } {\bf 12}, 2027 (2013); arXiv: 1109.0325.

\smallskip\noindent [8] D. Anguita, S. Ridella, F. Rivieccion, R. Zunino,
{\it Neural Networks} {\bf 16}, 763-770 (2003).

\smallskip\noindent [9] E. A\"imeur, G. Brassard, S. Gambs, 
Machine learning in a quantum world,
In Luc Lamontagne and Mario Marchand, editors, 
{\it Advances in Artificial Intelligence}, volume 4013 of 
Lecture Notes in Computer Science, page 431. Springer, Berlin/Heidelberg, 2006.

\smallskip\noindent [10] V. Giovannetti, 
S. Lloyd, L. Maccone,  {\it Phys.Rev.Lett.} {\bf 100}, 
160501 (2008); arXiv: 0708.1879.

\smallskip\noindent [11] V. Giovannetti, 
S. Lloyd, L. Maccone, {\it Phys.Rev.A} {\bf 78}, 
052310 (2008); arXiv: 0807.4994.

\smallskip\noindent [12] F. De Martini, V. Giovannetti, S. Lloyd, L. Maccone, 
E. Nagali, L. Sansoni, F. Sciarrino, {\it Phys. Rev. A } {\bf 80}, 
010302(R) (2009); arXiv: 0902.0222.

\smallskip\noindent [13] I. Chiorescu, N. Groll, S. Bertaina, T. Mori, S.
Miyashita, {\it Phys. Rev. B} {\bf 82}, 024413 (2010).

\smallskip\noindent [14] D.I. Schuster, A. P. Sears, E. Ginossar, L. DiCarlo, 
L. Frunzio, J. J. L. Morton, H. Wu, G. A. D. Briggs, B. B. Buckley, 
D. D. Awschalom, R. J. Schoelkopf,  
{\it Phys. Rev. Lett.} {\bf 105}, 140501 (2010).

\smallskip\noindent [15] Y. Kubo, F. R. Ong, P. Bertet, D. Vion, V. Jacques, 
D. Zheng, A. Dréau, J.-F. Roch, A. Auffeves, F. Jelezko, J. Wrachtrup, 
M. F. Barthe, P. Bergonzo, D. Esteve, 
{\it Phys. Rev. Lett.} {\bf 105}, 140502 (2010).

\smallskip\noindent [16] H. Wu, R.E. George, J.H. Wesenberg, 
K. Mølmer, D.I. Schuster, R.J. Schoelkopf, K.M. Itoh, 
A. Ardavan, J.J.L. Morton, G.A.D. Briggs, 
{\it Phys. Rev. Lett.} {\bf 105}, 140503 (2010).

\smallskip\noindent [17] M.S. Nielsen, I.L. Chuang, {\it Quantum
computation and quantum information}, Cambridge University Press,
Cambridge, 2000.



\smallskip\noindent [18] A.W. Harrow, A. Hassidim, S. Lloyd, 
{\it Phys. Rev. Lett.} {\bf 15}, 150502 (2009); 
arXiv: 0811.3171.

\smallskip\noindent [19] S. Aaronson, `BQP and the polynomial
hierarchy,'  arXiv:0910.4698.

\smallskip\noindent [20] L. Grover, T. Rudolpha `Creating superpositions
that correspond to efficiently integrable probability distributions,' 
arXiv: quant-ph/0208112.

\smallskip\noindent [21] P. Kaye, M. Mosca, in Proceedings of the 
International Conference on Quantum Information, Rochester, New York, 
2001; arXiv: quant-ph/0407102.

\smallskip\noindent [22] A. N. Soklakov, R. Schack, {\it Phys. Rev. A}
{\bf 73}, 012307 (2006).


\smallskip\noindent [23] S. Garnerone, P. Zanardi, D.A. Lidar,
{\it Phys. Rev. Lett.} {\bf 108}, 230506 (2012): arXiv: 1109.6546.

\smallskip\noindent [24] S. Lloyd, J.-J.E. Slotine, 
{\it Phys. Rev. A} {\bf 6201}, 2307 (2000); arXiv: quant-ph/9905064.












\vskip 1cm

\noindent{\it Supplementary material:} 

Here we present an
adiabatic algorithm for constructing a quantum state
$$ |\chi\rangle = (1/\sqrt M)\sum_{j}  |c_j\rangle |j\rangle
=  (1/\sqrt M)\sum_{c,j\in c}  |c\rangle |j\rangle
\eqno(S1)$$
that contains the output of the unsupervised $k$-means clustering
algorithm in quantum form.  This state contains a uniform superposition
of all the vectors, each assigned to its appropriate cluster,
and can be sampled to 
provide information about which states are in the same
or in different clusters.
For the quantum clustering algorithm, proceed as in the original
Lloyd's algorithm, but 
express all means in quantum superposition.
At the first step, select $k$ vectors with labels $i_c$ as 
initial seeds for each
of the clusters.  These may be chosen at random, or in a way
that maximizes the average distance between them, as in $k$-means$^{++}$.
Then re-cluster.  We show by induction that
the re-clustering can be performed efficiently by the quantum
adiabatic algorithm.  

For the first step, begin with the state
$${1\over  \sqrt{Mk}} \sum_{c'j}
|c'\rangle |j\rangle \big({1\over\sqrt k} \sum_c |c\rangle |i_c\rangle
\big)^{\otimes d}.\eqno(S2)$$
The multiple copies of the seed state ${1\over\sqrt k} \sum_c |c\rangle 
|i_c\rangle$ combined with the distance evaluation techniques
given in the paper allow one to
evaluate the distances $|\vec v_j - \vec v_{i_{c'}}|^2$ in
the $c'j$ component of the superposition, and to apply 
the phase $e^{-i\Delta t |\vec v_j - \vec v_{i_{c'}}|^2}$.
This is equivalent to applying the Hamiltonian
$$H_1 = \sum_{c'j} |\vec v_j - \vec v_{i_{c'}}|^2 |c'\rangle\langle c'|
\otimes |j\rangle\langle j|,\eqno(S3)$$
Now perform the adiabatic algorithm with the initial
Hamiltonian  
$H_0 = 1-|\phi\rangle \langle \phi|$, where
$|\phi\rangle = (1/\sqrt k) \sum_{c'} |c'\rangle$, 
adiatically deforming to the Hamiltonian $H_1$.
The time it takes to perform the adiabatic algorithm accurately
will be evaluated below.
The result is the first-order clustering state
$$|\psi_1\rangle = { 1\over \sqrt M} 
\sum_{c, j\in c} |c\rangle |j\rangle,\eqno(S4)$$
where each $j$ is associated with the $c$ with the closest
seed vector $i_c$.    By constructing multiple copies
of this state, one can also construct the individual
cluster states $|\phi^c_1\rangle = (1/\sqrt M_c) \sum_{j\in c} |j\rangle$
and estimate the number of states $M_c$ in the $c$'th cluster.
 
Now continue.  At the next re-clustering step,
assume that $d$ copies of the state $|\psi_1\rangle$ 
are made available from the previous step.  
The ability to construct the individual cluster states $|\phi_1^c\rangle$
together with the ability to perform the distance evaluation as
in the paper allows to
evaluate the average distance between $\vec v_j$ and the
mean of cluster $c$, $|\vec v_j - (1/M_c)\sum_{k\in c'} \vec v_k|^2 
= |\vec v_j - \bar v_c|^2 $.  This ability in turn allows
us to apply a phase  
$e^{-i|\vec v_j - \bar v_{c'}|^2  \delta t}$ to each component
$ |c'\rangle  |j\rangle$
of the superposition,  which is equivalent to applying
the Hamiltonian
$$ H_f = \sum_{c'j}  |\vec v_j - \bar v_{c'}|^2 
 |c'\rangle\langle c'|
\otimes |j\rangle\langle j|\otimes I^{\otimes d}.\eqno(S5)$$ 

Now,
perform the adiabatic algorithm,
starting with the state 
$$ {1\over  \sqrt{Mk}} \sum_{c',j}
|c'\rangle |j\rangle |\psi_1\rangle^{\otimes d} 
,\eqno(S6)$$
with initial Hamiltonian $1-|\phi\rangle \langle \phi|$, where
$|\phi\rangle = (1/\sqrt k) \sum_{c'} |c'\rangle$, 
and gradually deform to the final Hamiltonian $H_f$,
rotating the $|c'\rangle$ to associate each cluster label $c'$
with the set of $j$'s that should be assigned to $c'$.
We obtain the final state
$$ 
\bigg(~ {1\over\sqrt M}\sum_{c', j\in c'} |c'\rangle   
 |j\rangle ~ \bigg)
|\psi_1\rangle^{\otimes d} = |\psi_2\rangle |\psi_1\rangle^{\otimes d}
.\eqno(S7)$$  
That is,
the adiabatic algorithm can be used to assign states to clusters
in the next step of the quantum Lloyd's algorithm.

Repeating $d$ times to create $d$ copies, one can now iterate
this quantum adiabatic algorithm to create a quantum superposition
of the cluster assignments at each step.  
Continue the reassignment until the cluster assignment state
is unchanged (which can be verified, e.g., using a swap test).
Since Lloyd's 
algorithm typically converges after a small number of steps,
we rapidly arrive at the clustering state
$|\chi\rangle = (1/\sqrt{M})\sum_{c,j\in c}
|c\rangle |j\in c\rangle$.
The resulting $k$-means clustered quantum state $|\chi\rangle$
contains the final optimized $k$-means clusters in quantum
superposition and can be sampled
to obtain information about the contents of the individual clusters. 
  
To calculate the scaling of finding the clustering state, note first that 
each distance evaluation is essentially a weak measurement [24]
that perturbs the clustered state 
$|\psi_\ell\rangle^{\otimes d}$
at the previous level by an
amount $< d\sqrt d \delta^2$ (measured by fidelity),
where $\delta$ is the accuracy of the distance evaluation.
Accordingly, as long as the desired accuracy is
$\delta> 1/d^{2/3}$, $d$ copies of the next cluster assignment
state can be created from the $d$ copies of the previous
cluster assignment state.

To evaluate the time that the adiabatic algorithm takes note that
adiabatic part of the algorithm acts only on the $c'$ cluster labels,
and that the overlap squared between the initial state of each step
(S6) and the final state (S7) is $O(1/k)$.  Accordingly, the
time per step that the algorithm requires is no greater than
$O(k\log kMN)$ (and could be as small as 
$O(\log kMN)$ if the minimum gap during the adiabatic stage
is $O(1)$).  As Lloyd's algorithm typically converges after
a relatively small number of steps, our estimate for the overall
algorithm to construct the clustering state $|\chi\rangle$
is $O(k\log kMN)$.

\vfill\eject\end

Weak measurements [24] made on these 
$d$ copies allows the approximate evaluation of the
number of vectors $M_c$ in the $c$'th cluster,
as long as none of these $M_c$ are too small -- that is,
we require $M_c = O(M/k)$, so that each cluster contains
roughly the same number of vectors.
We will see below that $d$ can be made sufficiently large
to leave these $d$
copies of $|\psi_1\rangle$ essentially unchanged during the evaluation
of the $M_c$. 
By a set of conditional operations as above, create the state 
$${1\over  \sqrt{Mk}} \sum_{c'j}
|c'\rangle |j\rangle \bigg( {1\over \sqrt{2M} } \sum_c \sqrt{M_c} |c\rangle 
\big(~ |0\rangle |j\rangle + 
{1\over \sqrt{M_c}} |1\rangle \sum_{i\in c} |i\rangle~ \big) 
\bigg)^{\otimes d},\eqno(S5)$$  
where $M_c$ is the number of vectors in the $c$'th cluster.